\documentclass{nic-series}
\begin{document}
\title{Freezing and Collapse of Flexible Polymers}

\author{Thomas Vogel \and Michael Bachmann \and Wolfhard Janke}
\authortoc{Th. Vogel, M. Bachmann, W. Janke}

\institute{Institut f\"ur Theoretische Physik and Centre for Theoretical Sciences (NTZ), Universit\"at Leipzig,\\ 
Postfach 100\,920, D-04009 Leipzig, Germany\\
         \email{\{vogel, bachmann, janke\}@itp.uni-leipzig.de}}

\maketitle

\begin{abstracts}
	We analyze the freezing and collapse transition of a
	simple model for flexible polymer chains on simple cubic and
	face-centered cubic lattices by means of sophisticated
	chain-growth methods. In contrast to bond-fluctuation polymer
	models in certain parameter ranges, where these two
	conformational transitions were found to merge in the
	thermodynamic limit, we conclude from our results that the two
	transitions remain well-separated in the limit of infinite
	chain lengths. The reason for this qualitatively distinct
	behavior is presumably due to the ultrashort attractive
	interaction range in the lattice models considered here.
\end{abstracts}

\section{Introduction}

	It is well known that single homopolymer chains undergo a
	structural coil--globule transition at the so-called
	$\Theta$-point. Much theoretical, experimental, and
	algorithmic work was, and still is, spent to localize that
	point for various homopolymer models.
	In principle, there is no longer any difficulty to investigate
	lattice models up to very long chain lengths. A~well working
	technique to deal with the problem is the (n)PERM
	algorithm.~\cite{vogel_grassb97pre,vogel_grassb02jcp}

	Since a relatively short time, thanks to generalized-ensemble versions of PERM,~\cite{vogel_bachm04jcp,vogel_prellb04prl} it
	is also possible to investigate the temperature range far
	below the $\Theta$-temperature, where another
	transition called ground-state--globule, liquid--solid
	(crystallization) or freezing
	transition can occur.~\cite{vogel_bachm04jcp,vogel_rampaulbind07pre}
	A recent study of the bond-fluctuation model with respect to
	these different transitions for example showed, that the
	crystallization and the coil-globule transitions may, but
	generally do not, coincide in the thermodynamic limit,
	depending on the interaction
	range.~\cite{vogel_rampaulbind07pre}

	In this work, motivated by above mentioned studies, we will
	report results for the structural transitions of Interacting
	Self-Avoiding Walks (ISAW) on the simple cubic (sc) and
	face-centered cubic (fcc) lattice, the simplest model for
	flexible, interacting polymers.~\cite{vogel_vogel07pre}

\section{Model and Methods}
\label{vogel_sec_mod_meths}

	In the ISAW model, a local attractive interaction
	between non-bonded nearest neighbors is assumed. The total
	energy of an ISAW is given by $E=-n$, where $n$ is the number of such contacts.
	To simulate the model, we use the nPERM algorithm for the
	study of the $\Theta$-transition of very long chains and a
	generalized-ensemble version for the simulation of the
	low-temperature behavior of short polymers.
	The PERM method is a chain-growth algorithm based
	on the Rosenbluth method.~\cite{vogel_rosen55jcp} It includes population control by
	pruning or enriching populations during the growth, depending
	on threshold weights.~\cite{vogel_grassb97pre,vogel_grassb02jcp,vogel_bachm04jcp,vogel_prellb04prl}

\section{Results and Discussion}
\label{vogel_sec_results}

	For investigating structural transitions of the polymer model, we
	calculate the specific heat and analyze its peak structure. It
	is expected that even for polymers of finite length, peaks of
	fluctuating quantities signalize conformational activity.

\begin{figure}[h]
\begin{center}
\includegraphics[width=.495\textwidth,bb=105 453 476 669]{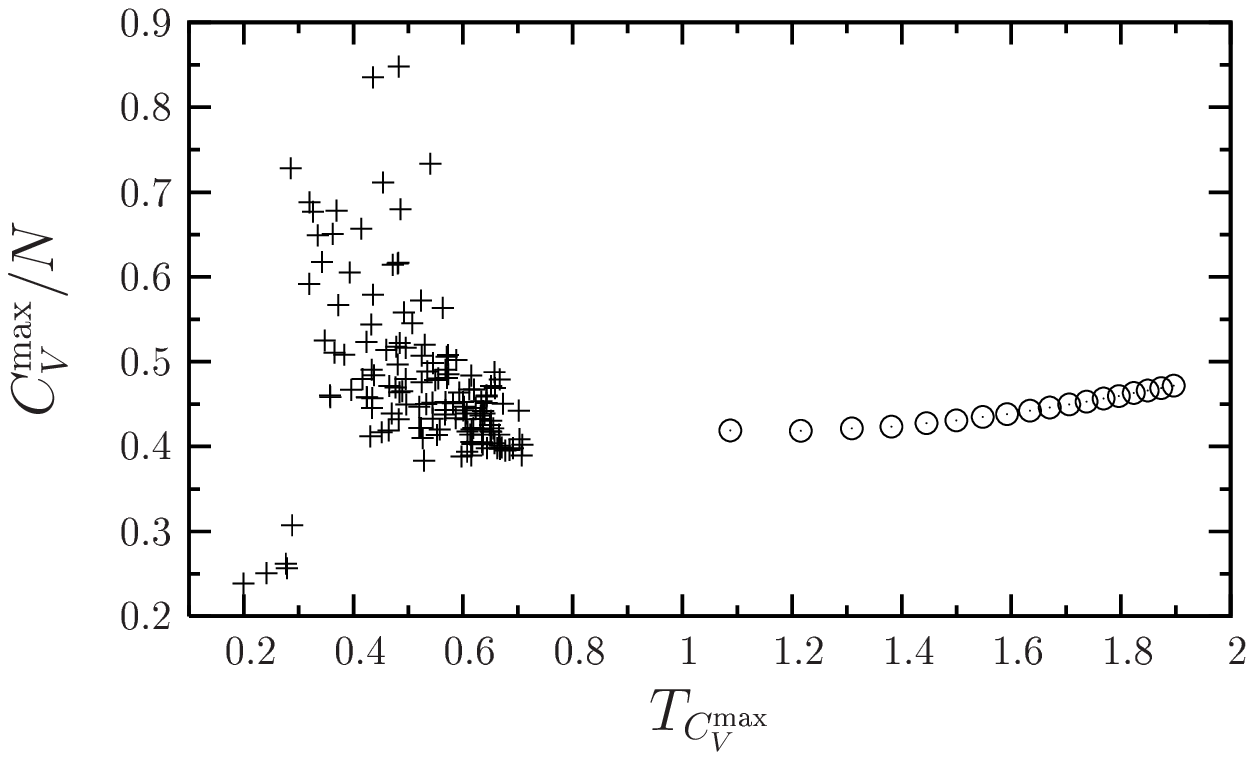}
\includegraphics[width=.485\textwidth,bb=92 450 461 669]{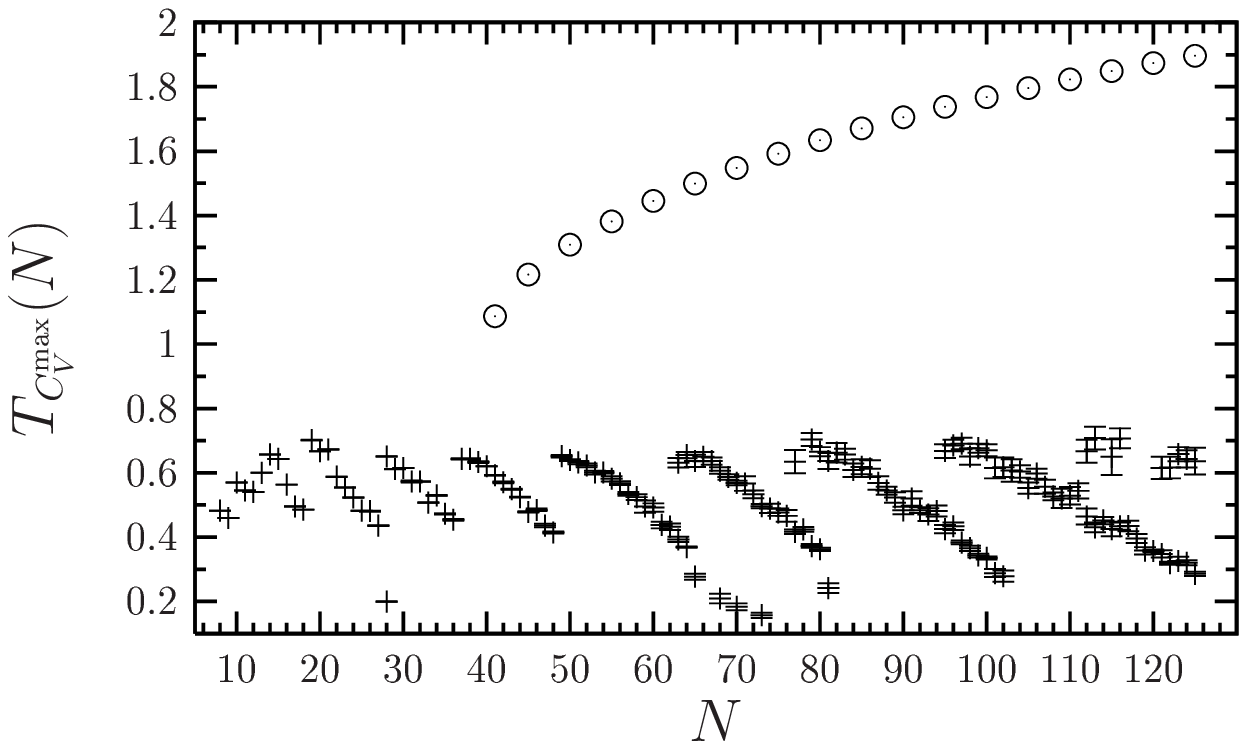}
\caption{\label{vogel_figure1}
Left: Map of specific-heat maxima for several chain lengths. Circles
($\odot$) symbolize the peaks (if any) identified as signals of the
collapse ($T_{C_V^{\rm max}}>1$). The low-temperature peaks ($+$)
belong to the excitation/freezing transitions ($T_{C_V^{\rm
max}}<0.8$). Right: Collapse and excitation/freezing
peak temperatures of the same specific-heat peaks.
}
\end{center}\quad\\[-2.8\baselineskip]
\end{figure}

	Figure~\ref{vogel_figure1}\,(left) shows specific-heat peaks
	of polymers on the sc lattice with
	lengths \hbox{$8\leq N\leq125$}. We see that there is
	no uniform scaling behavior of the peaks as it
	was found for the bond-fluctuation
	model.~\cite{vogel_rampaulbind05eurolett} A first view does not show
	any regularity at all regarding the low-temperature peaks.
	However, the dependence of the peak temperatures on the chain
	length exhibits more systematics, see
	Fig.~\ref{vogel_figure1}\,(right).

{
\noindent
	The freezing-transition temperatures show a sawtooth-like
	behavior which is due to optimal monomer alignment to the
	underlying lattice. At the lowest peak temperatures, we find
	chains with very compact ground states which are arranged as
	cubes or compact cuboids, respectively, e.g., for $N=27, 36, 48$, etc.
	The corresponding chains have an energy gap of $\Delta E=2$
	between the ground state and the first excited state. For the
	sc lattice, this can easily be explained. The first excited
	state can be constructed by removing a monomer from the corner
	of the compact state (breaking 3 contacts) and placing it
	somewhere at the surface (gaining 1 contact). All these chains
	have a very pronounced freezing-transition peak (not shown).
	On the other hand, at the other side of the ``teeth'', the
	respective chains with one more monomer reside. Here, the
	formerly pronounced low-temperature peak becomes very weak.
}

\begin{figure}[t]
\begin{center}
\includegraphics[width=0.4\textwidth]{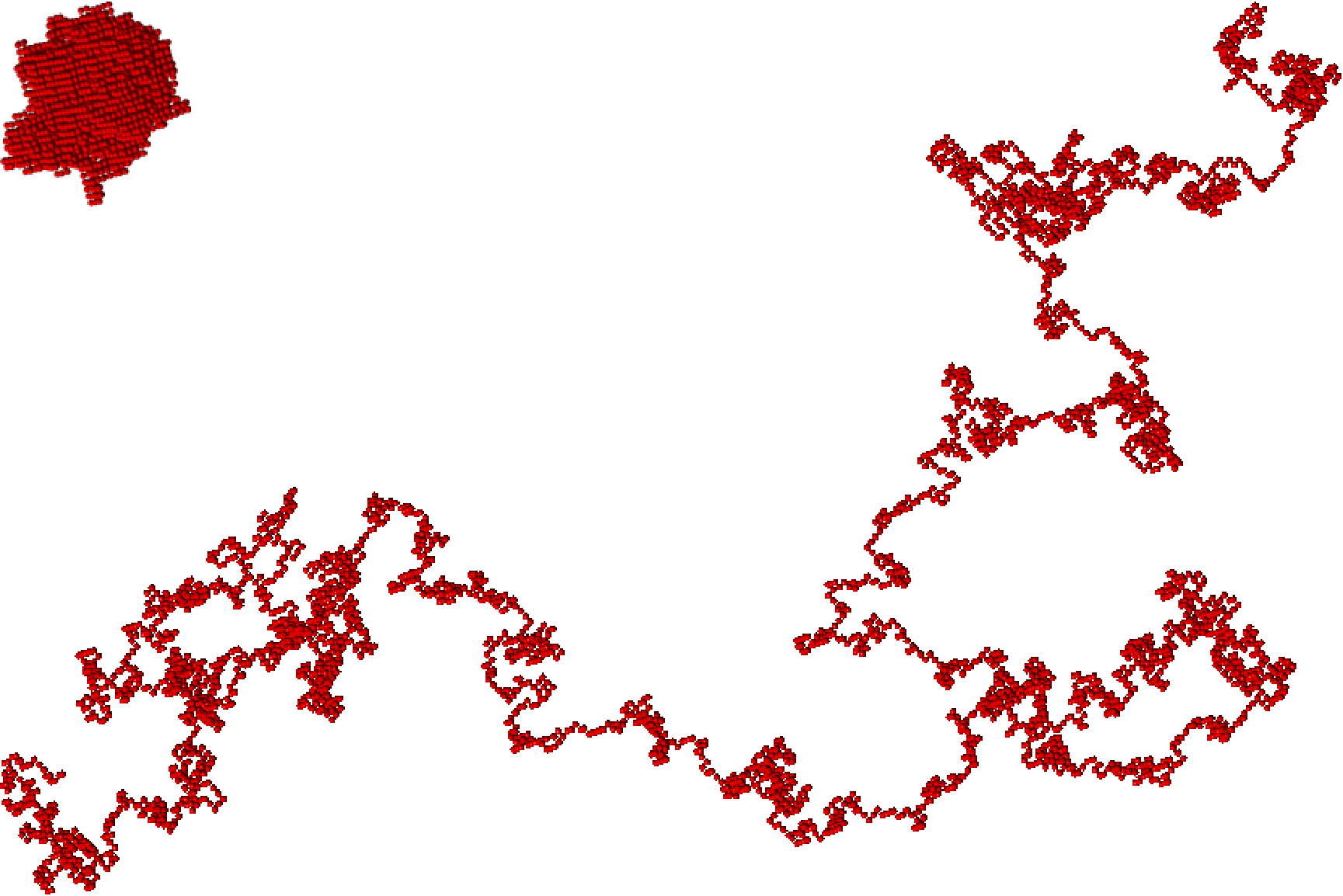}
\makebox[0.03\textwidth]{}
\includegraphics[width=0.51\textwidth,bb=106 444 466 663,clip]{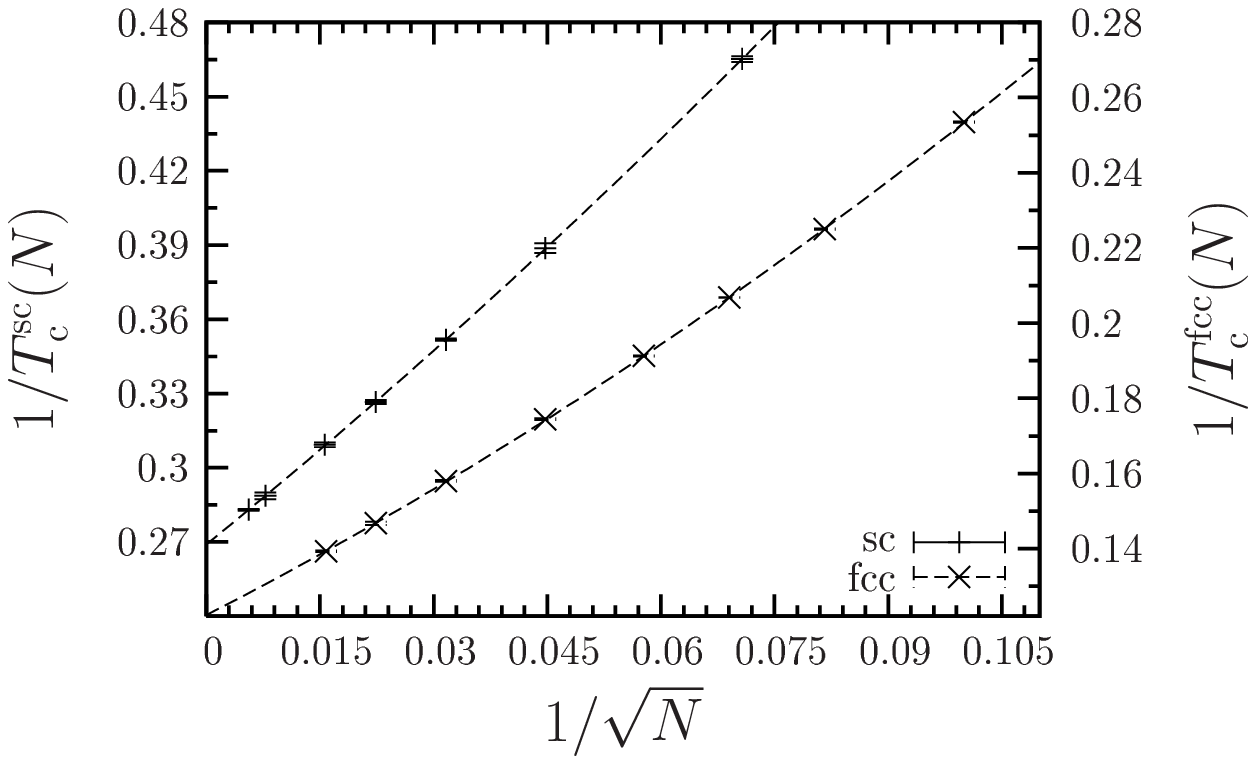}
\caption{\label{vogel_figure2}
	Left: Two typical conformations of a~4096mer on a sc
	lattice at different temperatures \hbox{$T\ll T_\Theta$};
	$E = -5367$ and $T \gg T_\Theta$;
	$E = -1054$. Right: Inverse collapse temperatures for
	several chain lengths on sc ($N\le 32\,000$) and fcc lattices
	($N\le 4\,000$). Drawn lines are fits according to
	Eq.~(\ref{vogel_eq1}). Error bars are shown, but may not be
	visible.}
\end{center}
\end{figure}

	Let us briefly say some words on the $\Theta$-point. Despite
	of very precise measurements, the nature of this transition is
	not yet completely understood, considering for example
	predicted logarithmic corrections~\cite{vogel_duplant86eurolett}
	which could not be resolved so far in numerical data.
	Figure~\ref{vogel_figure2} shows an illustrating picture of
	the studied objects (left) and data of transition temperatures
	for different chain lengths as well as fits to the data (right).
	There are several approaches to extrapolate $T_\Theta$ by
	fitting $T_\text{c}(N)$. We used here the mean-field like fit~\cite{vogel_vogel07pre}
\begin{equation}
\label{vogel_eq1}
\frac{1}{T_c(N)}-\frac{1}{T_\Theta}=\frac{{a}_1}{\sqrt{N}}+\frac{{a}_2}{N},
\end{equation}
	which was found to
	be consistent with numerical data obtained in grandcanonical
	analyses of lattice homopolymers and the bond-fluctuation
	model.~\cite{vogel_wilding96jcp}

	Optimal fit parameters using the data in the intervals $200\!\le\!
	N\!\le\!32\,000$ (sc) and $100\!\le\!N\le\!4\,000$ (fcc) were found
	to be $T_\Theta^{\rm sc}\!=\!3.72(1)$, ${a}_1\!\approx\! 2.5$,
	and ${a}_2\!\approx\! 8.0$ (sc), and $T_\Theta^{\rm
	fcc}\!=\!8.18(2)$, ${a}_1\!\approx\!1.0$, and
	${a}_2\!\approx\! 5.5$ (fcc). This agrees very well with
	data published so far,~\cite{vogel_grassb97pre,vogel_kremer96confproc} but
	does unfortunately not solve the problem of correction
	terms either.
	A very detailed analysis including different fit ans\"atze
	and parameter estimates can be found in Ref.~\citen{vogel_vogel07pre}.

	To summarize, there exists a clear low-temperature freezing
	transition below the \mbox{$\Theta$-point}, which is strongly
	influenced by lattice restrictions. It is shifted with
	increasing chain length to lower temperatures
	($T_{C_V^\text{max}}(N)\approx0.4$) and jumps at chains with
	very compact ground states back to a value of
	$T_{C_V^\text{max}}(N)\approx0.6$. The temperature interval,
	in which the freezing peaks fluctuate, does not change
	when varying the chain length, whereas the finite-length
	$\Theta$-temperature does.

\section*{Acknowledgments}
We thank K.\ Binder, W.\ Paul, F.\ Rampf, and T.\ Strauch for helpful
discussions.  This work is partially supported by the DFG under Grant
No.\ JA 483/24-1/2, the Graduate School ''BuildMoNa'', the DFH-UFA PhD 
College CDFA-02-07 and by a computer-time grant of the NIC at
Forschungszentrum J\"ulich, under No.\ hlz11.

\end{document}